\documentclass[12pt]{article}

\usepackage{times}
\usepackage{graphicx}
\usepackage{gensymb}
\usepackage{setspace} 
\usepackage{url}
\usepackage{color}

\newcounter{lastnote}

\title{Network Analyses and Nervous System Disorders}

\author
{John D. Medaglia,$^{1}$ Danielle S. Bassett$^{2,3\ast}$\\
	\normalsize{$^{1}$ Department of Psychology, University of Pennsylvania}\\
	\normalsize{Philadelphia, PA, 19104, USA}\\
	\normalsize{$^{2}$ Department of Bioengineering, University of Pennsylvania}\\
	\normalsize{Philadelphia, PA, 19104, USA}\\
	\normalsize{$^{3}$ Department of Electrical and Systems Engineering, University of Pennsylvania}\\
	\normalsize{Philadelphia, PA, 19104, USA}\\
	\\
	\normalsize{$^\ast$ To whom correspondence should be addressed; E-mail:  dsb@seas.upenn.edu.}
}

\begin{document} 

\baselineskip24pt

\maketitle 

\clearpage
\newpage

\singlespacing
\section*{Summary}
Network analyses in nervous system disorders involves constructing and analyzing anatomical and functional brain networks from neuroimaging data to describe and predict the clinical syndromes that result from neuropathology. A network view of neurological disease and clinical syndromes facilitates accurate quantitative characterizations and mathematical models of complex nervous system disorders with relatively simple tools drawn from the field of graph theory. Networks are predominantly constructed from \emph{in vivo} data acquired using physiological and neuroimaging techniques at the macroscale of nervous system organization. Studies support the emerging view that neuropsychiatric and neurological disorders result from pathological processes that disrupt the brain’s economically wired small-world organization. The lens of network science offers theoretical insight into progressive neurodegeneration, neuropsychological dysfunction, and potential anatomical targets for interventions ranging from pharmacological agents to brain stimulation.
\vspace{10mm}

\textbf{Keywords:} Network science | Neurology | Graph Theory | Connectome | Neuroimaging
\newpage


\section*{Introduction}

Nervous system disorders are characterized by pathological processes in the brain and peripheral neural tissue that can either be genetic or acquired. These disorders affect mental function and behavior as a natural consequence of neural degeneration or damage. The processes and effects of nervous system disorders are studied by biologists, psychologists, and clinicians, and also increasingly by individuals with backgrounds in technical and computational fields. In the neurosciences, an increasing synthesis between neurological and computational sciences is redefining how the function and dysfunction of the nervous system is quantified and treated. One intuitive mathematical framework for such a synthesis is \emph{network science}, a relatively new interdisciplinary field that provides theoretical and mathematical rigor to the study of nervous system disorders. Here, we offer the reader an overview of key concepts, methods, and findings from the application of network science to the neurosciences broadly defined, and we offer a vision for the future contributions of network analysis to our understanding and treatment of nervous system disorders.

\section*{Why Networks?}

To predict, diagnose, and treat nervous system disorders, we must simultaneously develop appropriate theoretical perspectives and analytic techniques that can transform the ever-increasing body of neural data into fundamental understanding. Historically, such perspectives and techniques have focused on univariate descriptors of neural function, behavior, and clinical outcomes. These univariate approaches have been critically important, offering insights into the consequences of local neural degeneration and damage on symptomatology. Nevertheless, evidence suggests that incorporating information from multiple variables of interest -- either from brain or behavior -- may offer more accurate biomarkers for disease and therapeutic interventions. Recently developed multivariate and machine learning techniques can be used to combine separate types of univariate information into predictive models to diagnose and treat patients. 

While multivariate techniques for data analysis are important, not all of them offer true theories of nervous system disorders. This limitation stems in part from the fact that establishing robust theoretical links between pathology, clinical syndromes, and treatment requires that the nervous system be studied at the appropriate scale(s) with adequate representations of the system of interest. A fundamental limitation to univariate and many common multivariate techniques is that they do not always represent the complex relationships within and between neural systems, genes, and behaviors. Addressing this limitation is critical for the study, diagnosis, and treatment of nervous system disorders, which carry with them heavy burdens both to individuals and to society.

Network analysis is well-positioned to address this gap. Having a mathematically rigorous foundation in graph theory, network analysis can be used to represent the complex pathological processes observable in nervous system disorders, to map pathology to mental function and behavior, and to bridge basic neuroscience and translational practice. These capabilities appear particularly appropriate for the study of the brain due to its inherent multi-scale network architecture \cite{bassett2011understanding}. Moreover, the utility of network analysis in understanding nervous system disorders is bolstered by the fact that the underlying pathophysiology often affects distributed network systems across multiple spatio-temporal scales \cite{Stam2014}. A natural goal for network analysis in nervous system disorders is therefore to identify network descriptions of pathophysiological processes and to use that knowledge to develop appropriate clinical interventions. 

\section*{Network Analysis}

Mathematically, a network can be represented by the graph $G = \{V,E\}$, where $V$ is a set of nodes that describe the components of a system, and where $E$ is a set of edges that describe relationships between nodes. In a typical network analysis, the graph $G$ represents a system and can be used to study the system's structure and function. Graph theory is thus concise in the sense that it offers a simple mathematical representation of a complex system, and is flexible in the sense that any system built from components and pairwise interactions can be represented as a graph. As an aside, we note that a graph is a natural representation of a system characterized by pairwise relationships, but systems that display higher order relationships require other representations \cite{giusti2016two}.

The graph $G$ can be parsimoniously encoded in an adjacency matrix, $\mathbf{A}$, whose $(i,j)^{th}$ element represents the weight of the edge between node $i$ and node $j$. Importantly, a network can be either \emph{binary}, where the elements of $\mathbf{A}$ are either 0 or 1, or it can be \emph{weighted}, where the elements of $\mathbf{A}$ take values within a given range. In addition, a network can be either \emph{undirected}, where edges represent bidirectional connections leading to a symmetric adjacency matrix, or it can be \emph{directed}, where the edge from node $i$ to node $j$ may have a different weight than the edge from node $j$ to node $i$. 

Selecting which network representation to use for a given system is an important process, and one that requires careful thought \cite{rubinov2010complex}. For systems in which relationships between components can be quantified accurately and confidently, a weighted network representation is appropriate \cite{bassett2016small}. If relationships can only be characterized confidently as present or absent, then a binary representation is appropriate. An undirected network may be used when no meaningful sense of direction exists in the data, whereas a directed network is often preferable to accurately represent a system in which one node causally influences another node. 

In the context of the nervous system, network analysis can be applied to any data for which there are at least two nodes and at least one edge. Thus, network analysis could be applied to the interactions between genes, neurons, neuronal ensembles, cortical columns, or Brodmann areas, or it could be applied simultaneously across two or more of these spatial scales. In this entry, we focus predominantly on studies examining the network structure and function of large-scale neural units (areas) within the human nervous system. 

\subsection*{Types of Networks in the Nervous System}

\begin{figure}[ht!]
 	\centerline{\includegraphics[width=6in]{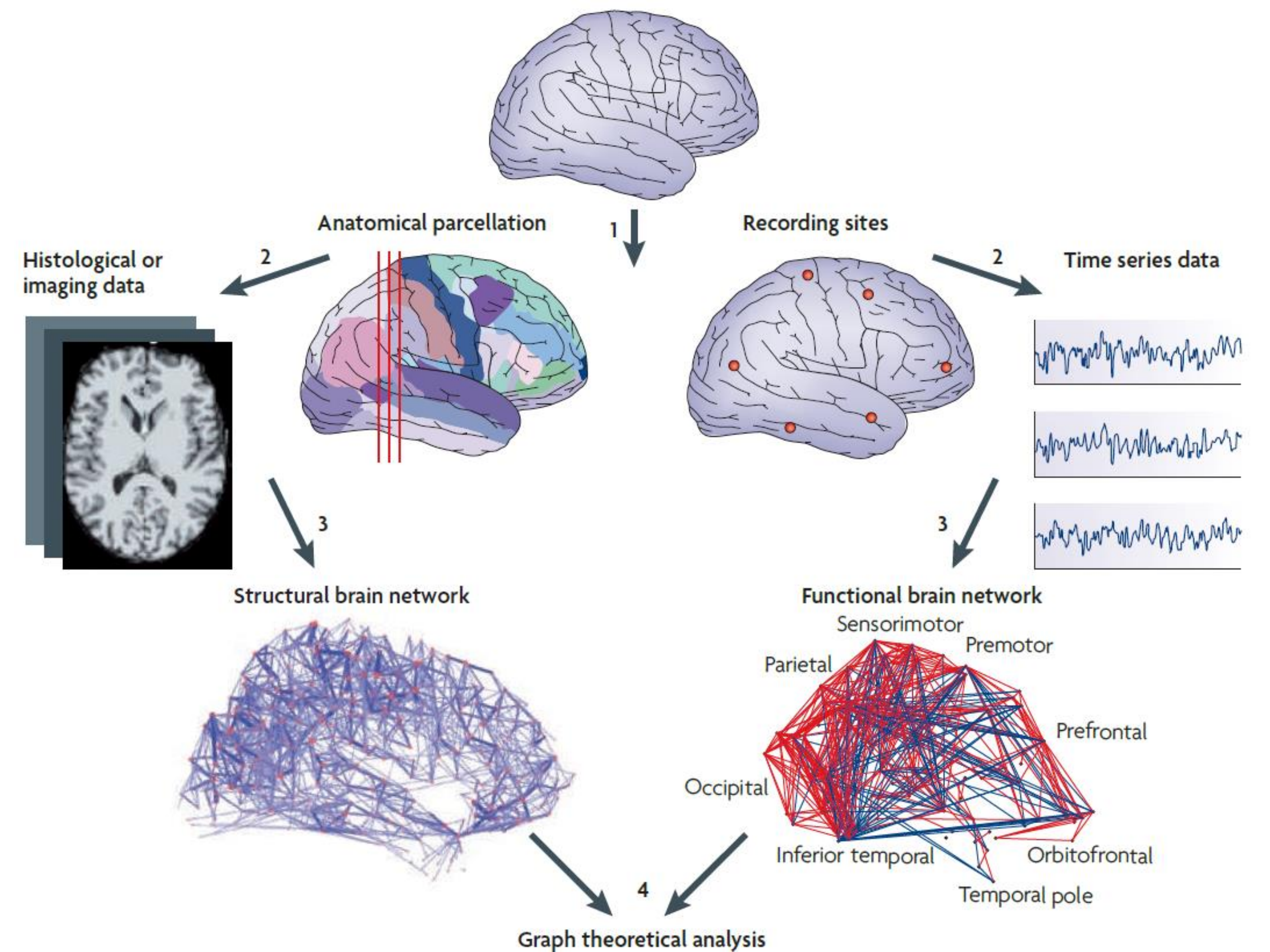}}
 	\caption{\textbf{Types of Networks.} The two most common categories of networks are \emph{functional} and \emph{structural}. Each category requires a different set of processing steps prior to the application of network analysis. \emph{Left}: Structural networks can be constructed by choosing a parcellation of the brain into regions of interest, applying that parcellation to anatomical imaging data, and estimating the structural relationships between regions of interest. \emph{Right}: Functional networks can be constructed by extracting signals from brain regions (which can also be parcels) and estimating the functional relationships between those regions using measures of statistical similarity or causality among pairs of time series. Figure reproduced with permission from \cite{Bullmore2009}.}\label{fig:networktypes}
 	\centering
 \end{figure}

Networks in nervous systems are typically categorized as \emph{structural} or \emph{functional} (Fig.~\ref{fig:networktypes}). In human neuroimaging, anatomical networks are often constructed from diffusion imaging data, which offers measurements of white matter microstructure. By applying sophisticated tractography algorithms to this data, one can estimate the probability with which one large-scale brain region connects to another via a white matter tract. These anatomical connections -- or physical pathways -- between all possible pairs of brain regions can be naturally represented in a graph, which can then provide information about fundamental organizational principles of brain architecture. 

Other sorts of structural features of brain tissue can also be used to construct graphs. For example, so-called \emph{morphometric} networks represent co-variation in gray matter thickness, gray matter density, surface area, surface curvature, or cortical thickness across subjects \cite{bassett2008,he2007,sanabria2010surface,ronan2012consistency}. Importantly, the architecture of a morphometric network must be interpreted in a fundamentally different manner than the architecture of a white matter network. An edge in a morphmetric network indicates that two brain areas display morphometric features that strongly co-vary over subjects, perhaps driven by mutually trophic effects \cite{Bloch2013}, or by common disease mechanisms \cite{sporns2010networks}.

Functional networks can be used to encode statistical similarities in the time series of regional activity, as measured by functional neuroimaging techniques including functional magnetic resonance imaging (fMRI), and electrophysiological techniques such as electroencephalography (EEG), magnetoencephalography (MEG), electrocorticography (ECoG), and multi-electrode recordings \cite{medaglia2015cognitive}. Common measures of association that quantify statistical similarities in regional time series include correlation, coherence, and mutual information. Here, network edges can represent communication, coordination, or shared influences. Edges can also represent causal relationships between areas, and their strength can be estimated by transfer entropy, dynamic causal modeling, or structural equation modeling.

 \newpage

\begin{figure}[h!]
 	\centerline{\includegraphics[width=3.5in]{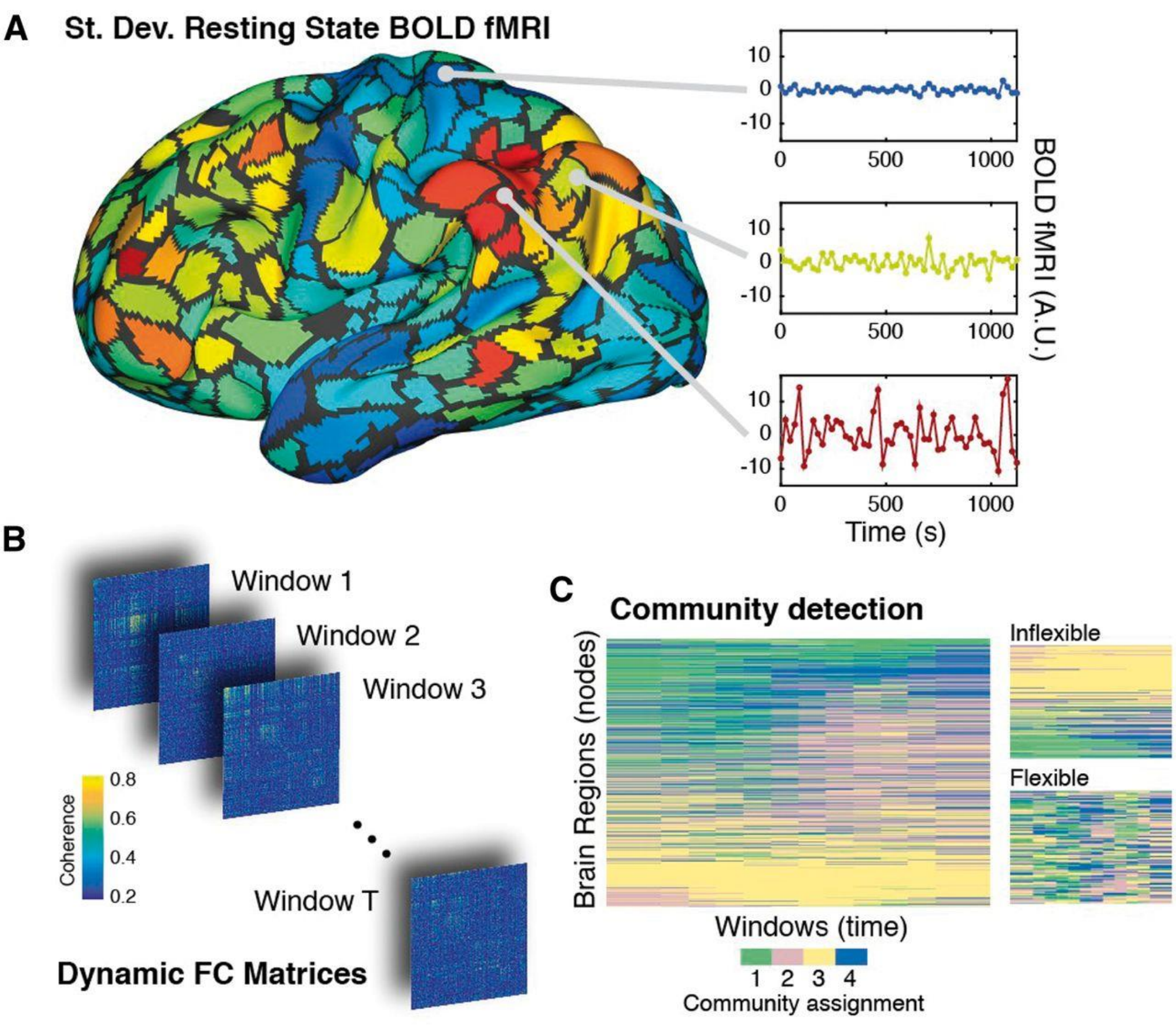}}
 	\caption{\textbf{Dynamic Networks.} An example time-varying network constructed from fMRI measurements extracted from different brain regions. \emph{(A)} The BOLD fMRI signal displays region-specific variability. \emph{(B)} Windowing time series and estimating the functional connectivity (FC) between pairs of regions reveals dynamic FC matrices. Each edge describes the statistical relationship (strength of connectivity) between two brain regions. The organization of FC matrices changes over time. \emph{(C)} Dynamic FC matrices can be used as an input to community detection algorithms to generate estimates of the network’s modular structure at each time point. Here, modules refer to collections of mutually-correlated brain regions that, as a group, are weakly correlated with the rest of the network. The dynamics of community structure can be characterized both in terms of individual brain regions and at the level of the whole network with the ``flexibility'' measure \cite{bassett2011dynamic}. Network flexibility indicates the extent to which brain regions change their community affiliation over time. Figure reproduced from \cite{mattar2016flexible} with permission.}\label{fig:dynamicnetwork}
 	\centering

 \end{figure}

In some cases, we might be interested in functional, morphometric, or structural networks that change over time \cite{calhoun2014chronnectome}: for example, on the order of milliseconds in the case of ECoG, seconds or minutes in the case of fMRI, and months or years in the case of diffusion imaging data. Efforts to address questions about how brain network architecture may change in a given time scale (or over multiple time scales) have led to the development of dynamic network analysis techniques \cite{bassett2011dynamic,doron2012dynamic}. One of the more common techniques for temporal networks that still derives from graph theory and network science is that of so-called \emph{multilayer} networks. A multilayer network is a network that is composed not just of a single adjacency matrix, but of multiple adjacency matrices linked to one another over time by identity links \cite{bassett2013robust}. While the tools are broadly applicable across data types, most studies to date have quantified the time-varying properties of functional networks at rest or during the performance of cognitively demanding tasks, and have examined both healthy subjects and patients with clinical syndromes \cite{braun2015human} (See Fig.~\ref{fig:dynamicnetwork}).

While it does seem natural to treat a region as a node in a brain network, recent work has also explored other construction methods in which nodes are used to represent other features of the data. For example, nodes can be patterns of activity at a fixed point in time, and edges can then encode similarities in the activity patterns of two different points in time. In another formulation, connections themselves can be treated as nodes, and similarities in the time series of connection weights can be treated as network edges \cite{davison2016individual}. These and other complementary construction methods can offer different insights into brain structure and function, and its alteration in nervous system disorders.

\subsection*{Characterizing Brain Networks Using Graph Statistics}

After carefully constructing the brain network that is appropriate for a given scientific question, one is faced with the challenge of quantitatively characterizing its structure and relating that characterization to its function. A relatively small number of graph statistics have been used frequently in brain network analysis to better understand and treat nervous system disorders. These include summary measures that describe entire networks as well as statistics that characterize the role of single regions or single edges \cite{rubinov2010complex}. Below, we address each of these types of statistics separately.

\paragraph{\emph{Node Statistics.}}

Several graph statistics characterize the role of a single node within the network. Arguably the simplest is the node \emph{degree}, which in binary networks can be defined as the number of edges connected to a node. In weighted networks, the \emph{weighted degree} (or \emph{strength}) is the sum of edge weights connected to a node. Node degree and strength represent the simplest form of \emph{centrality} in a network; nodes of high centrality are thought to be particularly influential on the network's function. Indeed, in brain networks, nodes with high degree are referred to as network hubs, and are thought to be critical for optimal information transmission and circuit-level computations \cite{sporns2010networks,betzel2016optimally}.

\begin{figure}[h!]
 	\centerline{\includegraphics[width=3.5in]{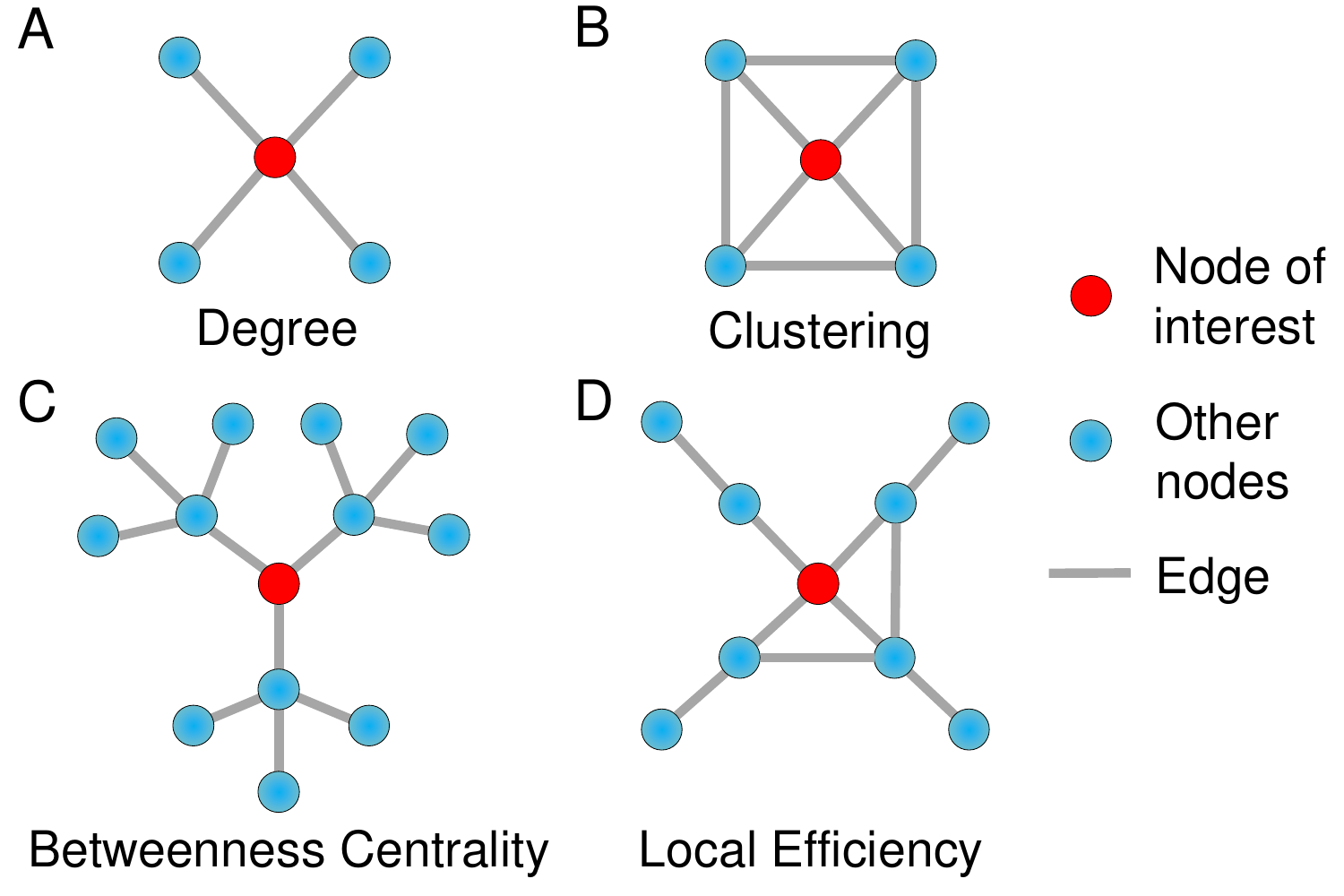}}
 	\caption{\textbf{Node Statistics.} The red node in each image denotes the node of interest. \emph{(A)} A node's degree is the number of edges emanating from a node. The red node in this graph has high degree, also known as degree centrality. \emph{(B)} The clustering coefficient represents the extent to which a node's neighbors are also connected to each other. The red node has high clustering coefficient because all of its neighbors are also connected to each other. \emph{(C)} Betweenness centrality quantifies the extent to which a node participates in shortest paths throughout the network; here, a shortest path is the path between node $i$ and node $j$ that traverses the fewest number of edges. The red node has high betweenness centrality. In this case, the red node also has high eigenvector centrality because it is connected three other nodes with high importance in the network. \emph{(D)} Local efficiency, which measures the shortest paths between all pairs of the neighboring nodes in the local subgraph. In this illustration, the local efficiency for the node of interest is low because few connections exist among the immediate neighbors of the node of interest.}\label{fig:nodestatistics}
 	\centering
 \end{figure}

Beyond degree and strength centrality, other centrality statistics include betweenness centrality, closeness centrality, and eigenvector centrality. Betweenness centrality quantifies the extent to which a node participates in shortest paths throughout the network; here, a shortest path is the path between node $i$ and node $j$ that traverses the fewest number of edges. Nodes with high betweenness are thought to be particularly influential brokers of information between different parts of the network. Closeness centrality quantifies the average shortest path between a given node and all other nodes in the graph; it is therefore commonly used as a measure of a node's ability to communicate quite broadly to the rest of the network. Eigenvector centrality uses the eigenspectrum of the adjacency matrix to quantify the influence of a node based on its connectedness with other high-scoring nodes in a network (See Fig. ~\ref{fig:nodestatistics}). Highly central nodes -- defined by any of these statistics -- are sometimes referred to as hubs given their predicted role in network function \cite{sporns2007identification}. 

Another useful metric to characterize a node's potential role in network function is the clustering coefficient, which describes how close its nearest neighbors are to being a clique (i.e., a completely connected subgraph). More specifically, the local clustering coefficient can be defined as the number of triangles in the network containing a node, divided by the number of connected triples containing that same node \cite{onnela2005intensity}. Intuitively, the clustering coefficient is a measure of the density of connections in a node's local neighborhood, and therefore it is often used to probe the potential of a node to participate in local information integration. A complementary notion is that of node ``efficiency'', which probes the organization of the edges among neighbors of a given node, thus offering a notion of the network's robustness to a node's removal \cite{latora2001efficient}.

\paragraph{\emph{Edge Statistics.}}

In addition to quantifying features of a node and its role in the network, one might also be interested in quantifying features of an edge. Perhaps the simplest statistic for an edge is its weight, which provides information about the strength of the relationship between two nodes. Moving beyond pairwise information, one can also compute something like the edge betweenness centrality, which measures the number of shortest paths between all possible pairs of nodes that pass through the edge of interest. Similar to the betweenness centrality of a node, the betweenness centrality of an edge is thought to represent its importance in efficient information transfer in networks. These are just a couple of examples of useful statistics for edges that can help one to understand the role of a node-node relationship in the broader network. 

\paragraph{\emph{Global Statistics.}}

While node and edge statistics are useful when one wishes to understand local structure in a network, more global statistics are important for characterizing over-arching organizational principles. Perhaps the simplest statistic is a network's density: the number of existing edges relative to the number of possible edges. Many natural networks exhibit sparse connectivity -- or low values of density -- due to the fact that edges are costly.  Arguably the most well-studied global characteristic of a network is the shape of its degree (or strength) distribution. Heavy-tailed degree distributions indicate the presence of unexpectedly large hubs. It is often interesting to ask whether these hubs preferentially connect to one another \cite{newman2002assortative}. An assortative network is one in which like-degree nodes tend to connect to one another, and a disassortative network is one in which unlike-degree nodes tend to connect to one another. In some networks, assortativity can be a marker of robustness: the removal of one high-degree node can be overcome by the interconnectedness of the others \cite{newman2002assortative,newman2003mixing}. 

Moreover, all networks can be characterized by average values of their nodal statistics: for example, the clustering coefficient of a network is equal to the average clustering coefficient of that network's nodes. High global clustering in a network indicates that nodes tend to be highly connected across all possible cliques of greater than two nodes. The \emph{characteristic} path length refers to the average shortest path between all pairs of nodes in the network. A short characteristic path length is thought to represent the potential for high integration across the network \cite{watts1998collective}. Global efficiency is a notion that is complementary to the characteristic path length, and can be calculated as the inverse of the harmonic mean of the shortest paths in the network \cite{latora2001efficient,Latora2003}. A network with a smaller characteristic path length has a higher global efficiency \cite{achard2007}. If one is interested in studying not just the shortest path but also longer paths or walks through the network, then one might wish to consider the statistic of network communicability \cite{estrada2008communicability}. From a practical point of view, it is important to note that measures of global clustering, path length, efficiency and others are heavily influenced by the network's density (or average strength), and it is therefore important to normalize graph density (strength) prior to making statistical inferences in the context of neuroimaging applications. 

\paragraph{\emph{Mesoscale Statistics.}}

In some cases, one wishes to ask whether a network displays a mesoscale organization: that is, a structure that is not easily characterized at either the local or global scales. A particularly important example of mesoscale organization is modular structure, which refers to the tendency for a network to have relatively dense connections between nodes located in a single module, and relatively sparse connections between nodes located in different modules. A second important example of mesoscale organization is core-periphery structure, which refers to the tendency for a network to have a core of densely interconnected nodes surrounded by a periphery of nodes that connect to the core but tend not to connect to other nodes in the periphery. Modularity is thought to offer a balance between the information integration and segregation capabilities of a network, while core-periphery structure can offer centralized processing.

\paragraph{\emph{Extending Graph Statistics to the Temporal Domain.}}

Recent efforts have begun to ask how we can extend graph statistics to the temporal domain. This question is motivated by the fact that brain network dynamics are inherently non-stationary, exhibiting temporally variable yet coordinated patterns of activity \cite{Kopell2014,calhoun2014chronnectome} that can be systemically altered in nervous system disorders. To quantify brain network (re)organization over time and its relationship to cognitive function, one can consider \emph{time-varying} network representations of neuroimaging data such as dynamic notions of modularity \cite{bassett2013robust} and core-periphery structure \cite{bassett2013task}. These and related approaches can be used to uncover fluctuations in statistics that occur at the node, edge, global, and mesoscale levels of brain network organization \cite{betzel2016multi}. Inherently dynamic statistics include flexibility \cite{bassett2011dynamic,mattar2016flexible} and promiscuity \cite{papadopoulos2016evolution}, which can be associated with symptomatology or neuropsychological test scores.

\subsection*{Using Graph Statistics}

Graph statistics are natural descriptors of network organization, and can be used to demonstrate (i) that the architecture of the brain network is unlike that expected in an appropriate null model network, (ii) that networks drawn from two or more cohorts display important similarities or differences, and (iii) that network architecture is non-trivially related to an external variable of interest, such as age or cognitive performance. In this section, we briefly discuss the utility of common graph statistics for these 3 general types of scientific questions.

\paragraph{\emph{Brain Networks \emph{versus} Null Model Networks.}}

When studying a brain network, one often wishes to demonstrate that the architecture observed in the real data is unlike that expected under a null hypothesis \cite{Bassett2011c,Bassett2012b}. Alternatively, one might also wish to demonstrate that the real data is \emph{like} a canonical graph model that has previously been demonstrated to afford important functional capabilities to the system. Thus, graph models are a critical part of brain network analysis \cite{Higham2008,Milenkovic2009,Rybarsch2012,Thorne2007,Kose2007}. 

\begin{figure}[h!]
	\centerline{\includegraphics[width=3.5in]{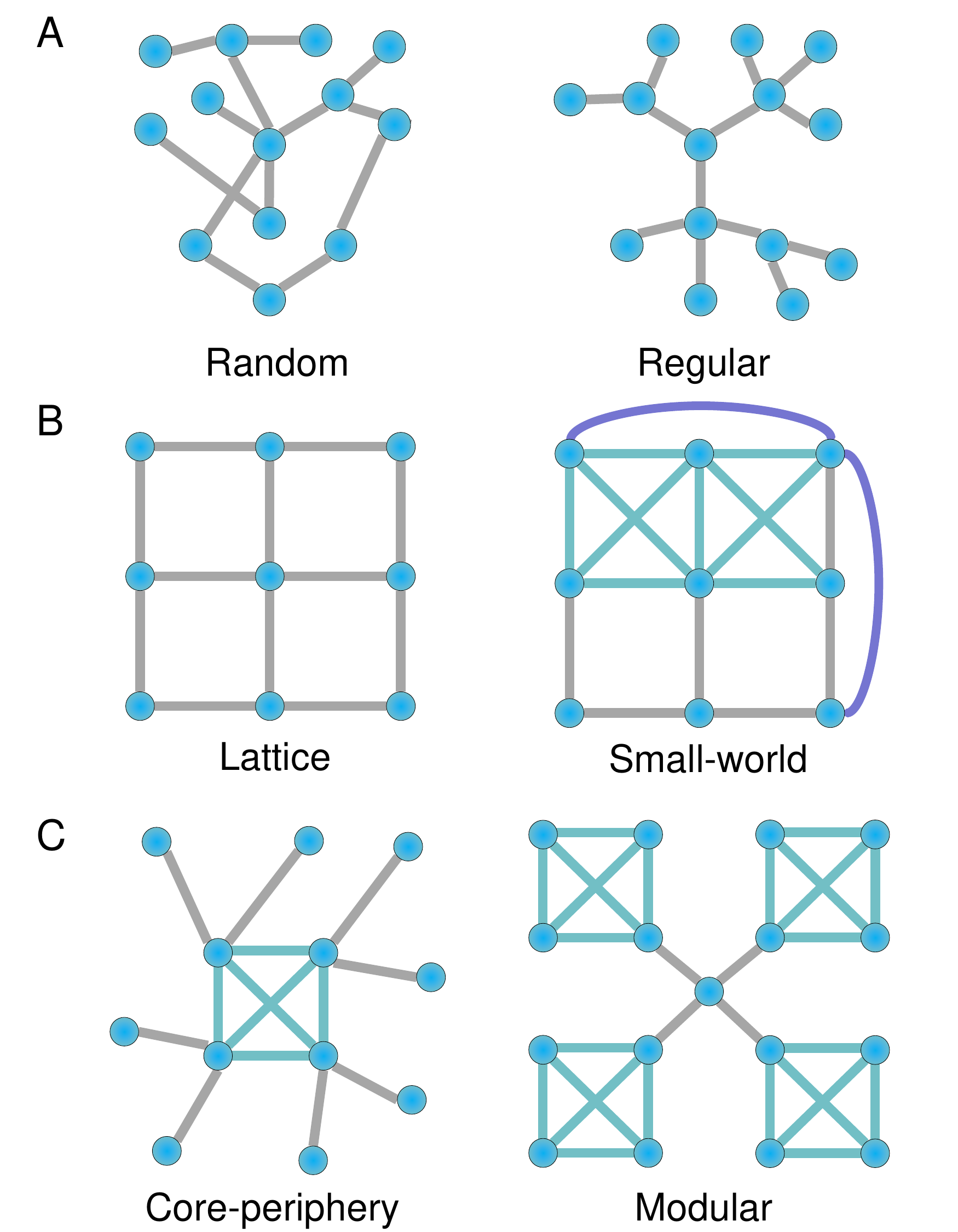}}
	\caption{\textbf{Canonical Architectures of Graphs.} \emph{(A)} \emph{Left}: A random network, in which each node has an equal probability of connecting to any other node, is characterized by a relatively short path length between any pair of nodes. \emph{Right}: A scale free network is built from a growing network model and is best characterized by a power-law degree distribution, indicating the presence of a very few, very high-degree hubs. \emph{(B)} \emph{Left}: A 2-dimensional regular lattice network in which each node is connected to its immediate neighbors on a square grid. \emph{Right}: A small-world network characterized by high clustering (similar to a regular lattice; teal edges) and short path length (similar to a random graph) induced by a few long-distance connections (purple edges). \emph{(C)} \emph{Left}: A core-periphery organization characterizing by a densely connected core and a sparsely connected periphery. \emph{Right}: A modular network with several highly segregated clusters.}
	\label{fig:globalstatistics}
	\centering
\end{figure}

An Erdos-Renyi graph is also called a random graph, and is characterized by the fact that nodes have an equal probability of connecting to one another \cite{Bollobas2001}. In contrast, a regular network is one in which all nodes have exactly $k$ neighbors. While a random graph has short path length, a regular graph has high clustering. Seminal work from Watts and Strogatz demonstrated that it is straightforward to show that a small-world graph is a natural intermediate architecture between these two models \cite{watts1998}. Small-world networks display high clustering similar to a regular graph but short path length similar to a random graph. Commonly observed in brain networks, this combination of architectural features can be estimated in real data \cite{muldoon2016small} and is thought to support a balance between information integration and segregation \cite{bassett2006small,bassett2016small}: the high clustering of the network allows for local processing to occur without being influenced by computations occurring at other nodes, whereas the short path length allows for ease of communication between nodes (or by extension modules) \cite{Stam2014}. Scale-free networks display a power-law degree distribution, while other networks exhibit a truncated power-law degree distribution, indicating the presence of a few highly connected hubs \cite{barabasi1999}. A different type of scaling is observed in fractal networks with hierarchically modular structure \cite{Sporns2006}. See Fig. \ref{fig:globalstatistics} for some example canonical graph structures.

Each of these graph models can be used as a null model (to prove architectural features that are unique to the brain) or a veritable model (to prove architectural features that are shared with the brain). These models fall into the category of models for static (time-invariant) networks. However, recent work has extended these notions to temporal (time-varying) networks. The most common temporal null models currently used in the literature are permutation based: one can permute time, node identity, and connection patterns \cite{Bassett2011b,Bassett2012b,Doron2012,Bassett2013}, for example, using a rewiring rule that maintains the underlying degree distribution \cite{Maslov2002}. Together, these null models are critical for statistical inference, and motivate the further development of other null models with increasingly realistic neurobiological structure. 

\paragraph{\emph{Group Summaries and Group Comparisons.}}

In addition to demonstrating that the architecture of a brain network is unlike that expected in an appropriate null model network, one might also wish to demonstrate that networks drawn from two or more cohorts display important similarities or differences. In some cases, addressing this question may begin with summarizing the network structure observed in a single cohort. While averaging networks across a group seems natural \cite{Achard2006,Song2009,Zuo2012}, this approach often fails to adequately reflect the topology of the individual networks in the group \cite{Simpson2012}. A promising alternative is the use of exponential random graph models \cite{Newman2010}, which can more accurately reflect the topological characteristics of the ensemble \cite{Simpson2011,Simpson2012}. Alternatively, one can perform analyses on single subjects, and then summarize those statistics over the group. For example, in the context of modular structure, one can construct a representative partition of network nodes into modules using consensus methods, which identify modules that are relatively consistent across a group of subjects or time points \cite{Lancichinetti2012,Bassett2012b}. 

After summarizing network architecture within a group, one might wish to compare network architecture between groups, such as between a patient sample and appropriately matched healthy controls. After computing a network statistic of interest in both samples, one can compare them using either parametric or non-parametric statistics. While parametric approaches such as $t$-tests, logistic regression, and analysis of variance can sometimes be useful, often the data violates the underlying assumptions of these approaches \cite{snijders1999non}. In contrast, non-parametric permutation tests that randomize the observed network data associations across groups can provide a more stringent statistical comparison.

\paragraph{\emph{Association and Prediction.}}

Beyond group differences, one also often wishes to demonstrate that network architecture is non-trivially related to an external variable of interest, such as age, cognitive performance, or clinical symptoms. Perhaps the most common method to relate a network statistic to a cognitive or behavioral variable is to compute the correlation coefficient between the two variables. If one is interested in the relationships between more than two variables, one could apply canonical correlation analysis, which can be used to relate many network features to clinical symptoms \cite{avants2010dementia}. Other multivariate techniques include multivariate (logistic) regression, mixed effects models, and causal inference techniques such as statistical mediation and structural equation modeling \cite{shipley2016cause}. These and related approaches enable the examination of cross-sectional assocations between network features and clinical symptoms, as well as longitudinal predictors of clinical symptomatology. The overarching goal is to clarify the network phenotypes associated with clinical syndromes, and to identify their underlying mechanisms.

\section*{Networks and the Healthy Nervous System}

\begin{figure}[h!]
 	\centerline{\includegraphics[width=6in]{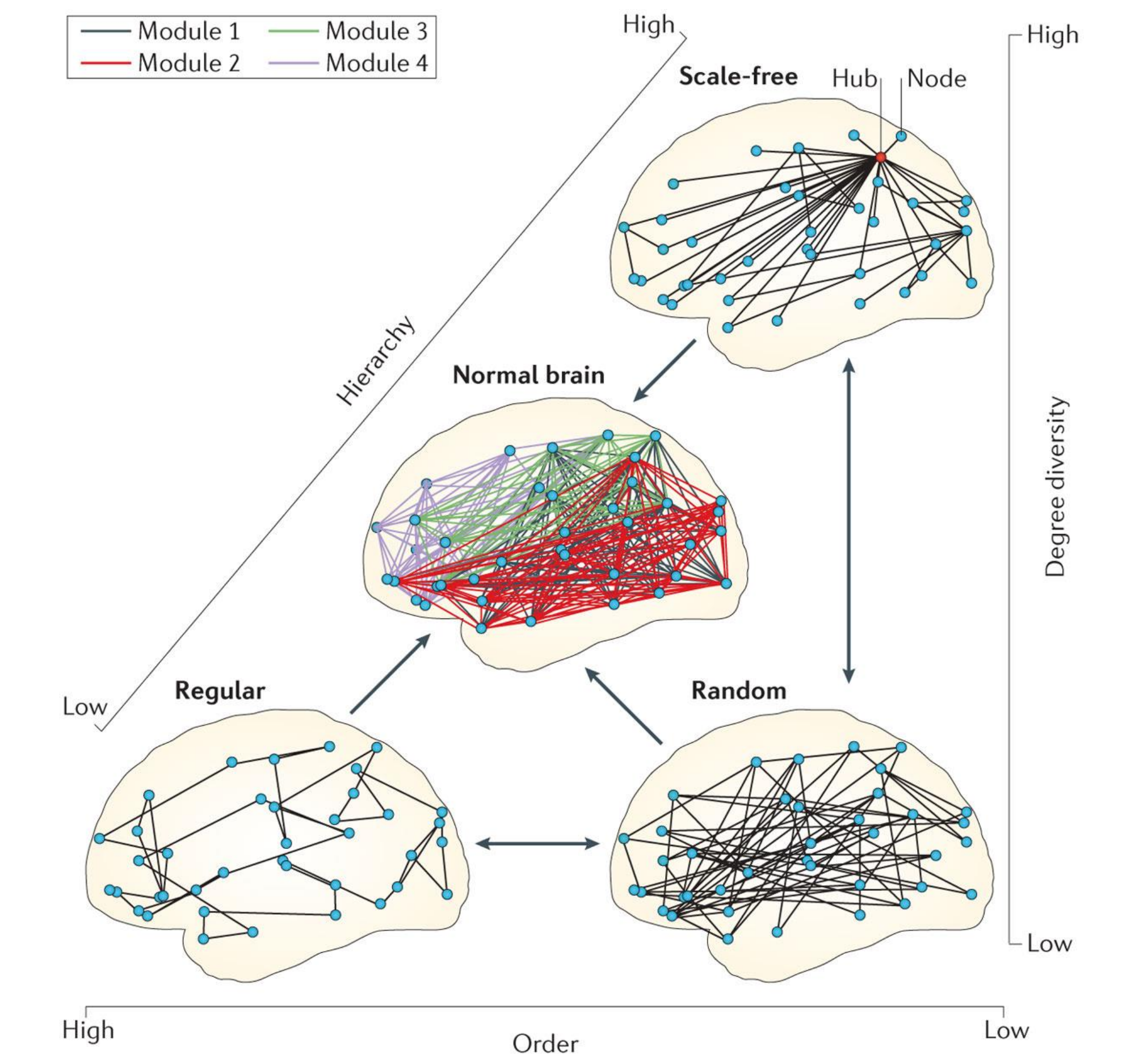}}
 	\caption{\textbf{Organization in healthy brain networks.} Healthy brain networks can be described as an intermediate between three extremes: a locally connected, highly ordered (`regular') network; a random network; and a scale-free network. Order is reflected in the high clustering of regular brain networks, whereas randomness (low order) is reflected in short path lengths. The scale-free component, characterized by a high diversity in node degree and high hierarchy, is indicated by the presence of highly connected ``hub'' nodes.  The composite of these attributes in normal brains results in a heirarchical, modular network. Figure reproduced with permission from \cite{Stam2014}.}
 	\label{fig:globalhealthyorganization}
 	\centering
 \end{figure}

To understand the alterations in brain networks that might accompany nervous system disorders, we must first understand the organization of brain networks in healthy subjects. There are several general organizational principles of brain network architecture that have been consistently observed across different cohorts, imaging modalities, and spatial scales of investigation. Perhaps the most historic is that of small-world organization \cite{bassett2006small,bassett2016small}, the simultaneous presence of both high clustering and short path length \cite{watts1998collective}, which is thought to support optimal information processing and transmission. This architecture is consistent with a heavy-tailed degree distribution indicating the presence of network hubs \cite{van2011rich}. At a mesoscale level, brain networks display heirarchical modular structure, with large modules being composed of smaller modules \cite{meunier2010modular,bassett2010efficient}. In both structural and functional networks, the large modules correspond to well-characterized cognitive systems including motor, auditory, visual, fronto-parietal, salience, and attention networks \cite{chen2008revealing,meunier2009age,power2011functional,yeo2011organization} (See Fig. \ref{fig:globalhealthyorganization}). Together, these organizational principles meet the evolutionary demand for information processing efficiency, flexibility, and robustness that is necessary for healthy cognitive function \cite{bullmore2012economy}

\newpage

\begin{figure}[h!]
	 	\centerline{\includegraphics[width=3.5in]{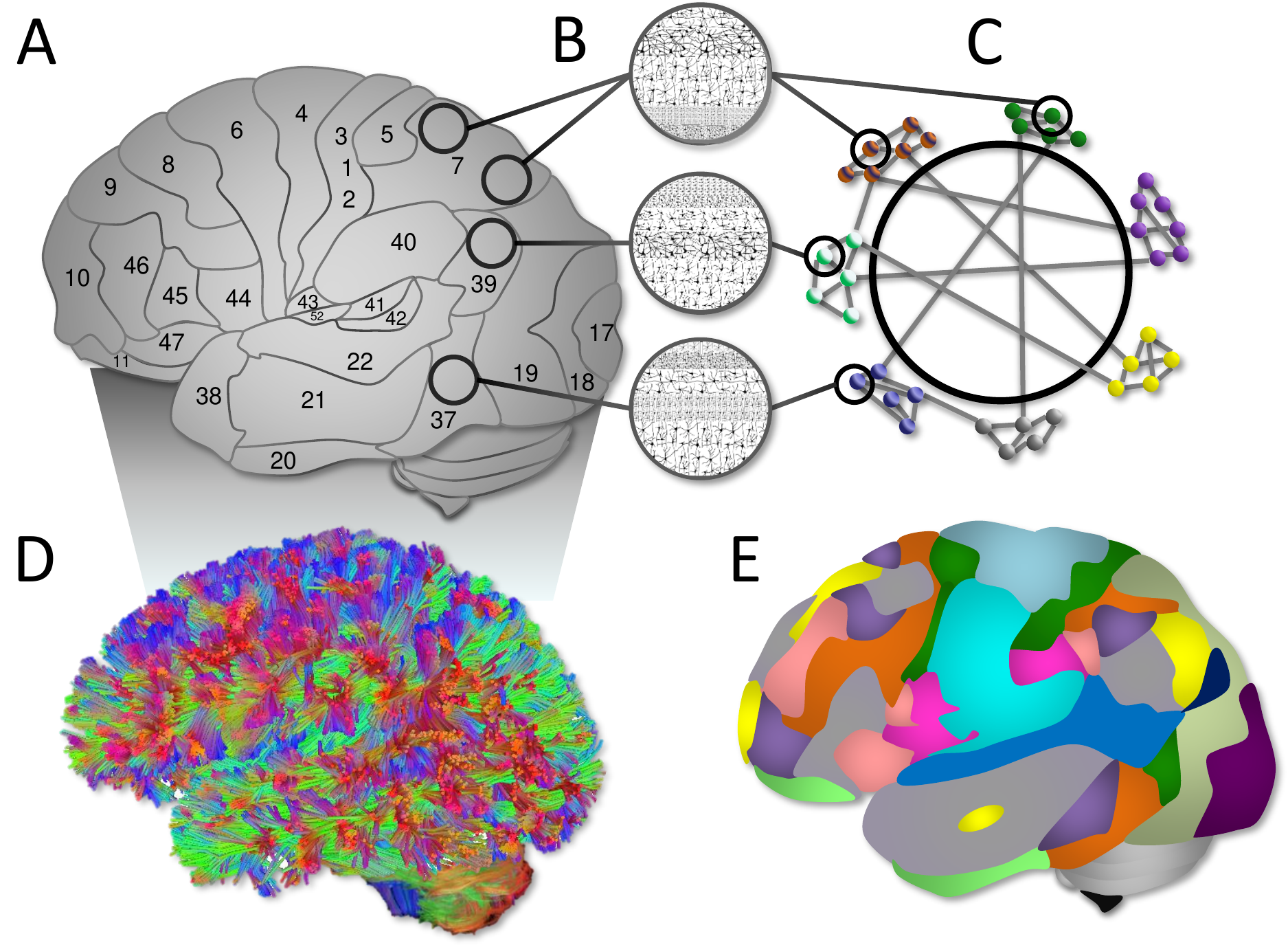}}
	 	\caption{\textbf{Organization in healthy brain networks.} \emph{(A)} Brain regions are organized into cytoarchitectonically distinct areas. \emph{(B)} Each area has a unique combination of structural properties that has specific implications for function. \emph{(C)} Areas can be represented as network nodes that have functional associations (edges) that extend beyond cytoarchitectonic boundaries. Groups of densely interconnected areas form network modules, which can have spatiotemporally varying connectivity patterns. \emph{(D)} Communication between brain regions is physically mediated by white matter pathways that can be estimated from diffusion imaging data. \emph{(E)} An example topography representing the modular organization of functional brain networks; note that modules can be physically dispersed, and therefore represent distributed communication between distinct types of computational resources. Figure reproduced and modified with permission from \cite{medaglia2015cognitive} and panel \emph{(E)} from \cite{yeo2011organization}.}
	 	\label{fig:nodestonetworks}
	 	\centering

	 \end{figure}

Exactly how different features of a network's organization support cognitive processes is a matter of ongoing research \cite{Sporns2014,medaglia2015cognitive}. Ultimately, the goal is to construct a validated theory that links the historic and ongoing efforts in careful neuroanatomy and physiology, to the recent perspective of the brain as a networked system or connectome, in a way that explains cognition and behavior. To be slightly more concrete, we might wish to understand (i) how the computations that occur in a single cytoarchitectonic region (originally characterized by Korbinian Brodmann in 1909 \cite{Brodmann1909}) are integrated with those of another cytoarchitectonic region via white matter tracts, (ii) how that integration leads to changes in the patterns of functional connectivity observed in functional neuroimaging data, and (iii) how local computations, transmission along white matter, and dynamic patterns of connectivity can support cognition (Fig.~\ref{fig:nodestonetworks}). While a full theory is far from defined, recent efforts address three major themes: the role of network ``hubs'', the architecture of intrinsic functional networks, and the existence of a complex heterarchy in the human brain.

\newpage

\subsection*{The Role of Hubs in the Healthy Nervous System}

\begin{figure}[h!]
 	\centerline{\includegraphics[width=3.5in]{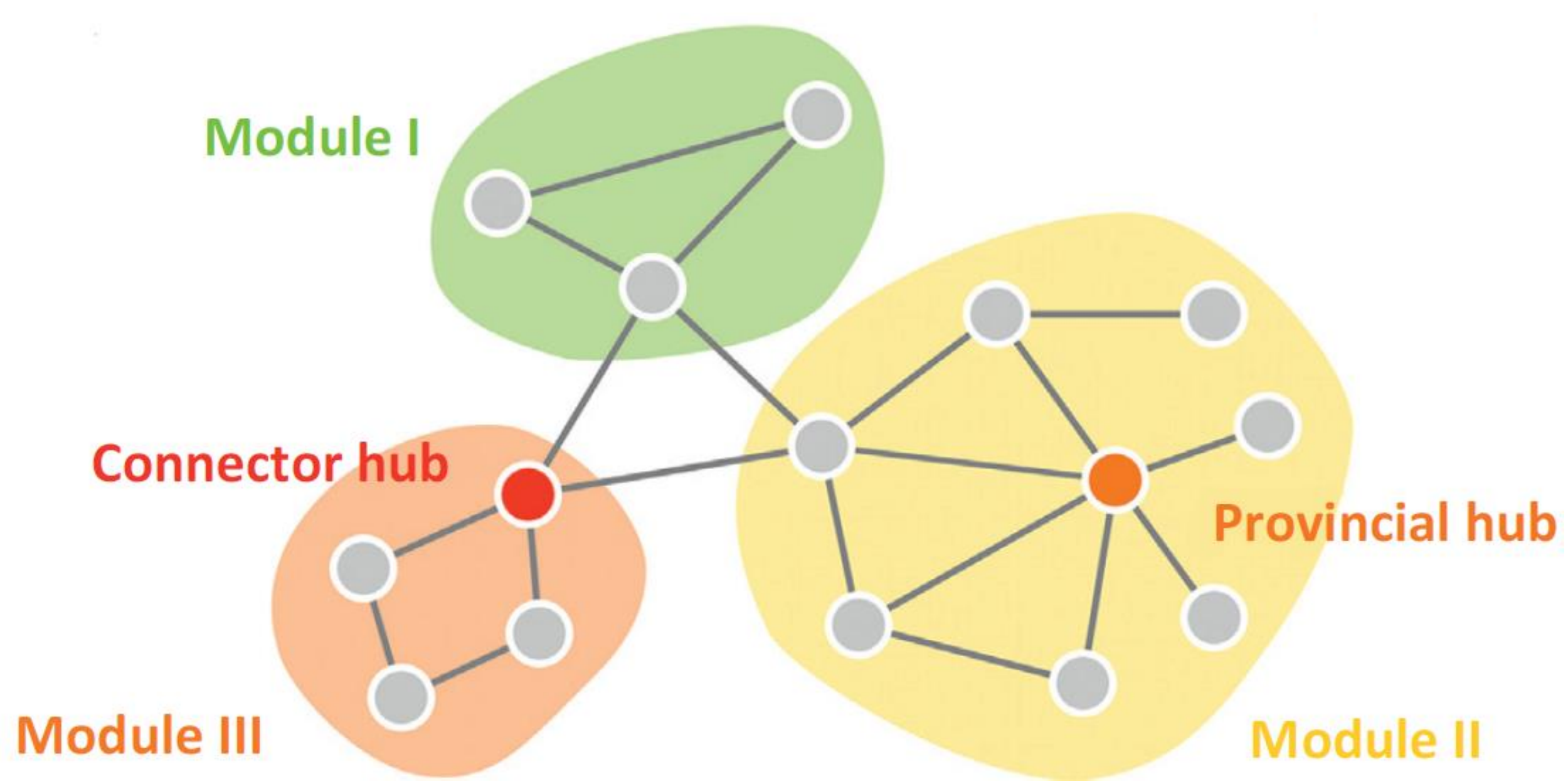}}
 	\caption{\textbf{Types of Network Hubs.} A module includes a subset of nodes of the network that show a relatively high level of within-module connectivity and a relatively low level of between-module connectivity. Provincial hubs are high-degree nodes that primarily connect to nodes in the same module. Note that this provincial hub is also a high-degree node in the entire network, which defines a more general notion of a hub. Connector hubs are high-degree nodes that show a diverse connectivity profile by connecting to several different modules within the network. Figure reproduced with permission from \cite{vandenHeuvel2013}.}
 	\label{fig:hubs}
 	\centering
 \end{figure}

Network hubs are interpreted as highly central nodes \cite{sporns2007brain}, but their mathematical definition is currently a bit nebulous. Most commonly, hubs are defined as nodes with high values of some centrality statistic \cite{sporns2007brain,buckner2009,Warren2014} (Fig.~\ref{fig:hubs}), putatively enabling them to regulate information processing by governing interregional communication along short paths through the network \cite{buckner2009}. In structural networks built from diffusion imaging data, hubs are located in the precuneus, anterior and posterior cingulate cortex, insular cortex, superior frontal cortex, temporal cortex, and lateral parietal cortex \cite{hagmann2008mapping,van2011rich,van2012high,van2013anatomical}, areas of the brain that are commonly thought of as transmodal or heteromodal areas involved in integrating processing across several cognitive modalities \cite{mesulam1998sensation}.  In functional networks, hubs are concentrated in the ventral and dorsal precuneus, posterior and anterior cingulate gyrus, ventromedial frontal cortex, and inferior parietal brain areas. Some of these regions overlap with the default mode system that displays high activation when subjects are simply resting \cite{raichle2001default}, while others overlap with the fronto-parietal system, and are known to adaptively reconfigure communication with different cognitive systems during task performance \cite{cole2013multi,cole2016activity}.

Hubs are suggested to be important for cognition because they are located along shortest paths in the nework, and therefore are likely to play a critical role in distributed patterns of communication. This location is evident both by their high degree, and by their tendency to connect to one another, forming a core or ``rich-club'' \cite{hagmann2008mapping,van2011rich,van2012high,van2013anatomical,van2011rich} that boosts the robustness of inter-hub communication and promotes efficient communication across the brain \cite{van2011rich}. Hubs are a cost-efficient solution to the problem of increasing network efficiency to support cogntive processes without requiring many metabolically expensive connections, and their existence is therefore thought to have been driven by evolutionary pressures \cite{van2013high}. Because of their central role in the network's topology, damage to brain hubs -- particularly those in the rich club \cite{van2011rich} -- is expected to be especially consequential for cognitive function in terms of the severity and pervasiveness of cognitive deficits \cite{warren2014network}.

\newpage

\subsection*{The Architecture of Intrinsic Brain Networks}

The first two decades of fMRI studies of human cognition largely focused on signal amplitude and connectivity that were elicited or evoked by the performance of a cognitively demanding task. However, recent evidence demonstrates that the brain also displays interesting network architecture even outside of the explicit requirements of a task. This baseline connectivity is often referred to as \emph{intrinsic} functional connectivity \cite{Seeley2007}, and is characterized by strong inter-regional correlations in BOLD magnitudes in the default mode network \cite{raichle2001default}: ventromedial frontal cortices, anterior and posterior cingulate, and inferior parietal areas. The default mode network can be consistently identified whether subjects are resting with eyes closed or eyes open, and -- interestingly -- becomes deactivated when a subject moves from rest to the performance of a cognitively demanding task \cite{raichle2001default}. 

In addition to the default mode network, intrinsic functional connectivity patterns also show strong inter-regional correlations in BOLD signal in other sets of brain areas. These sets are referred to as modules, networks, or cognitive systems in common parlance \cite{power2011functional}, and include motor, visual, auditory, memory, fronto-parietal, cingulo-opercular, salience, and attention systems. Regions within these systems activate together as a unit during the performance of cognitive tasks \cite{tavor2016task}, and they also communicate with one another across a variety of tasks \cite{cole2014intrinsic}, with slight variations in patterns of connectivity in different tasks \cite{cole2013multi}. Interestingly, the patterns in BOLD activation that are elicited by cognitively demanding tasks can be predicted by the time-averaged patterns of functional connectivity using a simplified activity "flow" model \cite{cole2016activity}, suggesting that fast-time-scale activity and and slow-time-scale connectivity are simply related to one another.

 \begin{figure}[h!]
 	\centerline{\includegraphics[width=3.5in]{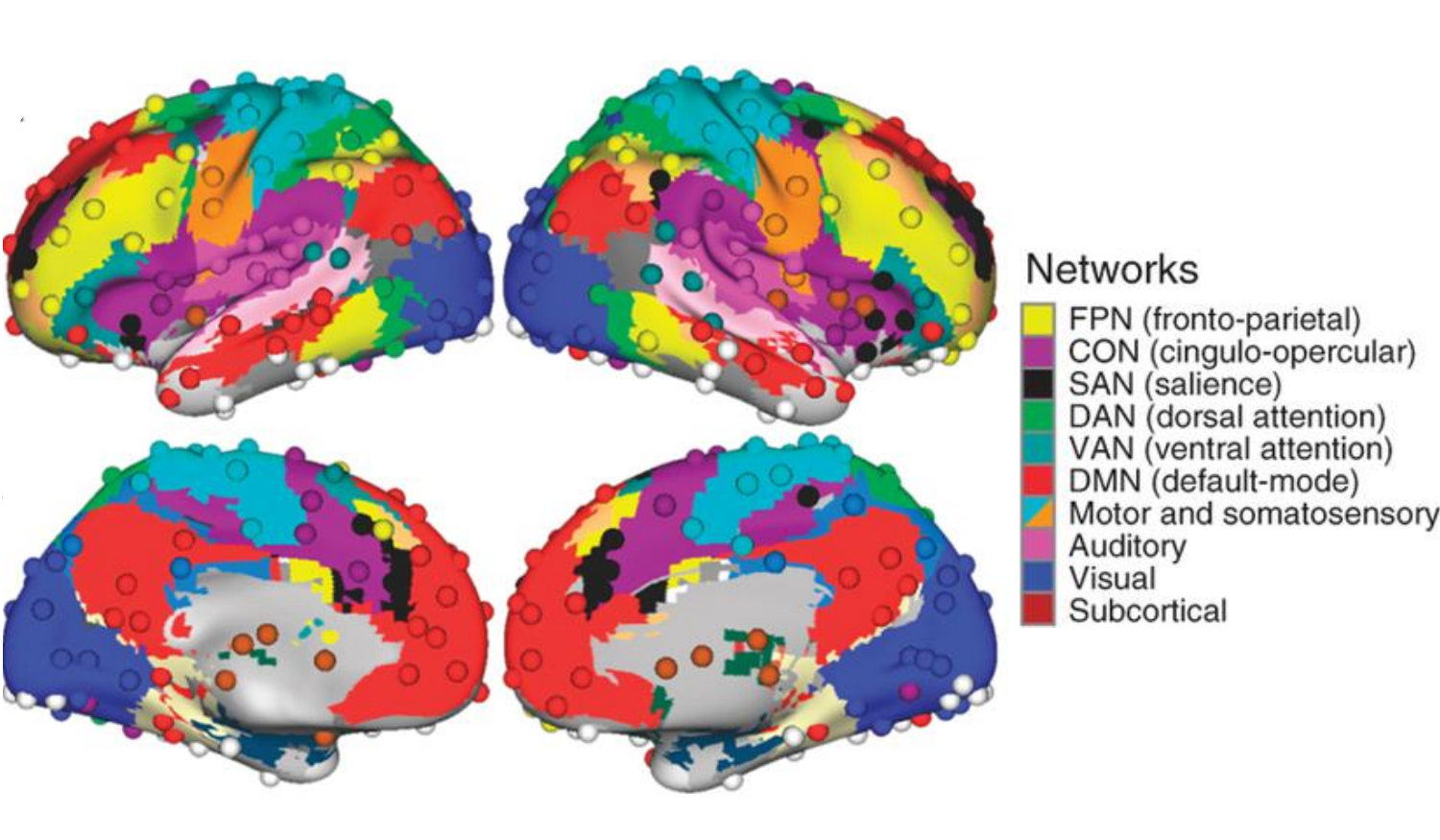}}
 	\caption{\textbf{Intrinsic networks in the human brain.} Intrinsic patterns of functional connectivity display strong inter-regional correlations in BOLD signal in sets of brain areas that are commonly referred to as modules, networks, or cognitive systems \cite{power2011functional}. These include the default mode, motor, visual, auditory, memory, fronto-parietal, cingulo-opercular, salience, and attention systems. Figure reproduced with permission from \cite{cole2013multi}.}
 	\label{fig:intrinsicnetworks}
 	\centering
 \end{figure}

\subsection*{Complex Heterarchy}

The combined presence of both network hubs and network modules supports the notion that there may be several different dimensions of organization in the brain that critically support the complexity of cognitive function. That is, it is not simply enough to list network hubs to understand cognition, as such a list would miss the fact that regions are grouped into systems that collectively support a cognitive process, such as decision-making, reward processing, or visual perception. Likewise, simply listing the modules in the network that support each process overlooks how those modules are integrated with one another, largely through the connectivity of network hubs within the brain's rich-club. Thus, these two features of network organization -- hubs and modules -- form orthogonal dimensions that support cognition.

The notion that different parts of a system can have different rankings of importance depending on the taxonomy used is not new in neuroscience. In fact, in 1945, McCulloch suggested that the brain is a complex heterarchy, characterized by multiple hierarchies defined in different ways \cite{McCulloch1945}. A full understanding of cognitive processes may therefore require a description of the hubs that are activated by the process, the modules that are formed or that change during the process, and the core-periphery structure that might support the process. Moreover, it may also require an understanding of how these notions of ranking or hierarchy map to the more traditional notions of higher-order cognitive systems being closer to the pinnacle of cognition than sensorimotor systems \cite{sepulcre2012,sepulcre2014} (although see \cite{persichetti2015value} for other perspectives on the presence of higher-order information in primary areas). Such a comprehensive description of cognitive function is an important area of active research in the field of network neuroscience.

\section*{Major Findings in Network Analysis for Nervous System Disorders.}

The emerging network neuroscience of healthy humans sets the stage for understanding nervous system disorders. In particular, neurological dysfunction can be thought of as the consequence of damage to brain regions and connections. It is reasonable to hypothesize that the disordered function characterstic of clinical syndromes can be linked to alterations in the brain's intrinsic connectivity, pathways from sensorimotor cortex to higher-order cognitive areas, network hubs, and the rich-club. To address this hypothesis, recent efforts have applied network analysis to neuroimaging data acquired in patients with nervous systems disorders, largely focusing on describing alterations in global, mesoscale, and local network organization. In the following sections, we highlight a few major findings from anatomical and functional network analysis in nervous system disorders. Given the already vast number of studies in this area, we pay particular attention to essential themes in clinical research, and we refer the reader to reviews for more specific information where relevant.

\subsection*{Network Disruption in Nervous System Disorders}

Studies aimed at detecting disruptions in the brain's functional and structural organization in nervous system disorders date back to the early 21st century. Here we highlight major results from network analysis in several of the most frequently examined neurological disorders and we refer the reader to \cite{Stam2014} for further details.

\paragraph{Neurodegenerative disorders.}

Neurodegenerative disorders are characterized by progressive loss of neural tissue due to insidious biological pathology that manifests in different locations in the brain, and across different spatial scales of organization. The clinical syndromes associated with neurodegenerative disorders often begin to be detected in adults. Examples include diseases with onsets late in life such as Alzheimer's disease, frontotemporal lobar degeneration, or Parkinson's disease, middle age onset such as in Huntington's disease, and variable onset amyotrophic lateral sclerosis. Each disorder is associated with complex symptom profiles that span both cognitive and motor domains. A major challenge in neurodegenerative disease is to articulate theories of functional loss, predict disease progression, and identify potential treatment mechanisms to preclude, halt, or reverse disease. 

Network-based studies of neurodegenerative disorders have been most prominent in Alzheimer's disease (AD) and frontotemporal lobar degeneration (FTLD). Studies constructing functional brain networks from electroencephalography (EEG), magnetoencephalography (MEG), and positron emission tomography (PET) acquired at rest have reported reduced clustering coefficients and local efficiency in AD in comparison to healthy controls \cite{dubbelink2014predicting,skidmore2011connectivity,agosta2013brain,heringa2014multiple,tijms2013single,reijmer2013disruption,de2009functional,stam2009graph,de2014effect,supekar2008network,brier2014functional,seo2013whole,lo2010diffusion}, although an opposite trend has been observed in fMRI data \cite{wang2013amnestic,he2008structural,yao2010abnormal,zhao2012disrupted,liu2012altered}. Node centrality is consistently decreased in AD in higher-ordered association areas, including the temporal lobe, medial parietal, posterior and anterior cingulate, and medial frontal areas \cite{dubbelink2013disrupted,agosta2013brain,de2012activity,lo2010diffusion,brier2014functional,seo2013whole,yao2010abnormal,minati2014widespread}. Moreover, the preference of hubs to connect to one another is decreased in  AD \cite{agosta2013brain,de2009functional} but increased in FTD \cite{agosta2013brain,de2009functional}. These central and high-degree areas also show the greatest relative amount of amyloid deposition in AD \cite{buckner2009cortical}, suggesting a possible link between amyloid pathology and the vulnerability of network hubs \cite{Stam2014}. 

While the trends for clustering and centrality are fairly consistent across studies, those for path length are less consistent. Increased path length in certain disorders has been reported in studies using group-level cortical thickness correlations, MRI tractography and fMRI \cite{agosta2013brain,he2008structural,lo2010diffusion,yao2010abnormal,liu2012altered}, potentially explaining the observed decrease in the AD brain's predicted ability to synchronize \cite{tahaei2012synchronizability}. However, even using a single imaging technology, such as fMRI, opposite changes in path length have been reported in the same disorder \cite{agosta2013brain,sanz2010loss,zhao2012disrupted,liu2012altered}. These discrepancies could be driven by differences in processing techniques \cite{Stam2014}, and therefore the question of whether functional differences in path length are pathognomonic to neurodegeneration remains open. 

Studies of mesoscale architecture have suggested that modular organization in the brain is disrupted in AD \cite{brier2014functional,de2012activity,chen2013modular}, with particular alterations in the parietal module. A recent MEG study showed that both the connections within this module and the connections between this module and other modules are decreased in AD \cite{de2012activity}. Notably, a pattern of module disconnection has also been observed in studies that used fMRI data \cite{brier2014functional,chen2013modular,cciftcci2011minimum}. Importantly, these findings suggest that at least some of the presenting clinical features in neurodegeneration are associated with disruption in information integration and segregation across specialized functional modules in the brain.


\paragraph{Multiple Sclerosis.}

Multiple sclerosis (MS) is a progressive demyelinating disorder characterized by inflammation and degeneration in myelin in the brain. MS can occur in isolated attacks or can display a more continuous progression. MS typically has an onset between 20 and 50 years of age and is associated with symptoms including double vision, unilateral blindness, muscle weakness, sensation deficits, and discoordination. MS is observed twice as commonly in women as men and lowers life expectancy by 5 to 10 years \cite{goldenberg2012multiple}.

Several studies have characterized altered functional network topology in MS. An MEG study showed increased functional connectivity in the theta, lower alpha and beta bands, and decreased functional connectivity in the upper alpha band in individuals with MS compared with healthy controls \cite{schoonheim2013functional}. In addition, changes in the normalized clustering coefficient in the lower alpha band were associated with impaired cognition. A second MEG study investigated networks in source space by using a connectivity measure that was unbiased by volume conduction \cite{tewarie2013cognitive}. Compared with healthy controls, individuals with MS had functional networks that were more integrated in the theta band and more segregated in the alpha and beta bands. A disruption of hierarchical network organization in the upper alpha band was associated with impaired cognition, whereas disturbed beta band connectivity of the default mode network was associated with both impaired cognition and motor deficits \cite{tewarie2013cognitive}. 

One large study provided evidence for a shift from global to local connectivity in cognitively impaired MS patients relative to unimpaired patients \cite{van2014graph}. Finally, a resting-state fMRI study found a correlation between decreased functional connectivity and cognitive impairment in male patients with MS but not in female patients \cite{schoonheim2013functional}. Thus, across functional studies, evidence suggests that cognitive impairments may result from disruptions in heirarchical and modular network organization, and these features may be specific to middle frequency ranges in the human brain.

Turning to structure and morphometry, we note that cortical thickness analyses have revealed altered small-world topology in MS that correlates with white matter pathology \cite{he2009impaired}. Diffusion tractography networks display reduced global efficiency and reduced local efficiency in the default mode, and in primary sensory and motor cortices \cite{shu2011diffusion}. These alterations occurred alongside increased node centrality in other brain regions such as the orbital frontal, superior frontal, middle frontal, and fusiform gyri.

Overall, MS has received a relatively less extensive network analysis focus in the literature. The links between cognition, specific network changes, interactions with sex, and frequency-specific changes have not been thoroughly established. In addition, some network statistics are found to be increased and others decreased in MS anatomy, and no intuitive relationship to cognitive decline has been identified. It is timely to invigorate efforts into examining functional disturbances in MS to identify mechanisms of cognitive decline.

\paragraph{Traumatic Brain Injury.}

Traumatic brain injury (TBI) can result from exogenous forces such as penetrating (``open'') or blunt force (``closed") trauma to the brain, or from endogenous hemorrhaging that results in stroke. Closed TBI often involves damage to neural bodies in gray matter depending on the location of trauma. TBI can also give rise to diffuse axonal injury across white matter tracts in the brain due to the stretching and shearing that results from forceful impacts and rotations of the cranium.

Anatomical studies of TBI reveal increased shortest path length and decreased global efficiency \cite{caeyenberghs2014altered}. In addition, hub regions are altered in TBI, including the reduction of hub scores for the left superior frontal gyrus, right superior parietal gyrus, and right postcentral gyrus and increases in hub scores for the left precentral gyrus, right supplementary motor area, and left putamen \cite{caeyenberghs2012graph}. Moreover, in a larger study, machine learning approaches identified that betweenness centrality and eigenvector centrality were associated with patient processing speed, executive function, and associative memory in a sample where these same statistics were reduced within network hubs \cite{fagerholm2015disconnection}. Collectively, these data suggest that global and local measures of information processing efficiency are selectively disrupted in TBI and contribute to cognitive deficits.

Most studies of functional network organization in TBI have used resting state fMRI. Early in recovery, TBI is associated with increases in functional connectivity that decrease in the first few months following injury \cite{hillary2015hyperconnectivity}. Later following TBI, the strength of long-distance connections decreases \cite{messe2013specific,nakamura2009resting}, increasing the network's path length and by extension disrupting the brain's small-world topology, which only partially recovers over time \cite{han2014disrupted,achard2012hubs}. Like AD, TBI is characterized by the selective damage of hub-like structures in the association cortices and the default mode network \cite{pandit2013traumatic,achard2012hubs}, and notably these changes are associated with impaired consciousness in TBI \cite{achard2012hubs}. Specifically, hubs are redistributed from the fusiform gyrus and precuneus to the angular gyrus in patients with TBI \cite{achard2012hubs}. TBI is also associated with changes in the modular structure of functional brain networks \cite{messe2013specific,han2014disrupted}. Individuals who sustained shock blast demonstrated a reduced participation coefficient \cite{han2014disrupted}, indicating selective disruption to intermodular communication. 

These studies suggest that functionally, TBI can result in a loss of (especially long-distance) functional connectivity, presumably as a result of the combined influences of initial diffuse axonal injury followed by neuroplastic processes. These network-level changes can have a direct impact on cognitive dysfunction. In a sample of individuals with stroke, focal damage to the inferior parietal lobe, which participates in several major functional brain systems, is strongly associated with pronounced global cognitive impairment \cite{Warren2014}, whereas damage to other high degree regions produced more circumscribed deficits. This promotes the important notion that a node's centrality in the network's topology can be indicative of that node's potential for catastrophic failure when damaged.

\paragraph{Epilepsy.}

Epilepsy is a seizure disorder that can be inherited or that can result from infection (meningitis), tumors, abnormal dysplastic tissue formation, or brain damage. Seizures can vary from brief and nearly undetectable to long periods of vigorous shaking. Epileptic seizures are often believed to have a focus from which seizures originate \cite{hader2013complications}. The mechanisms of epilepsy are unknown in the majority of cases, though known genetic mutations are linked to a small proportion of cases \cite{pandolfo2011genetics}. Across afflicted individuals, the onset of seizures is highly variable in severity and frequency over the lifespan, and the effect on quality of life can range from minimal to severity debilitating \cite{tellez2005long}.

The focus of network studies of epilepsy has largely been on functional networks, and particularly those derived from EEG and MEG recordings due to the fast timescale of seizure dynamics. Studies that have assessed network changes that occur during the ictal state have produced the most consistent results. The connectivity of the default mode network is diminished both in individuals with temporal lobe and generalized epilepsy \cite{douw2010epilepsy,douw2010functional}. In addition, an early study demonstrated high clustering and long path length during the ictal state \cite{ponten2007small}, consistent with intracranial studies reporting a shift towards a more regular network topology \cite{kramer2008emergent,schindler2008evolving,takahashi2012state}. Ictal network regularization is also observed in scalp EEG and MEG recordings \cite{ponten2009indications,gupta2011space}. 

Some studies show that EEG-and MEG-based functional brain networks in epileptics demonstrate abnormally regular networks during the \emph{interictal} state as well \cite{chavez2010functional,horstmann2010state}. One study suggested that in the interictal state, regularization might occur in the theta band, whereas alpha band networks are abnormally randomized \cite{quraan2013altered}. Other studies have also shown an association between excessive synchronization in the theta band and epilepsy \cite{clemens2013neurophysiology,douw2010epilepsy,douw2010functional}. In general, the interictal network topology in individuals with epilepsy seems to be shifted away from the small-world organization that is seen in healthy subjects towards an excessive regularity that is also observed during seizures.

Other studies also suggest the importance of hub-like structures, particularly in high-frequency ranges, for the spreading of seizure activity \cite{amini2010comparison,varotto2012epileptogenic}. Hub-like features in EEG recordings can predict whether children will develop epilepsy after an initial seizure-like event \cite{van2013improved}. Two studies compared the phase synchronization and node centrality of functional networks exhibiting high-frequency oscillations that are characteristic of epileptogenic tissue \cite{ibrahim2013neocortical,van2013high}. Counterintuitively, the number of oscillations was negatively correlated with node centrality in the theta band \cite{van2013high}. This finding suggests that areas with high frequency oscillations and hub nodes may differ, and could represent two different portions of the epileptogenic zone. Seizure state analyses suggest that seizures constitute a finite set of brain states that progress consistently, and that seizure onset zones are isolated from surrounding regions at seizure onset \cite{burns2014network}. During seizure progression, topographical and geometrical changes in network connectivity strengthen and tighten synchronization near seizure foci \cite{khambhati2015dynamic}. Notably, the epileptic zone's activity may be controlled by a regulatory network in the surrounding tissue: A simulated resection study in drug-resistant epilepsy revealed that antagonistic ``push-pull'' interactions between regions synchronize and desynchronize the network, potentially suggesting a candidate for therapeutic targets \cite{khambhati2016virtual}. 

Finally, a few recent studies have combined fMRI with anatomical MRI in the same subjects to study structure-function relationships in epilepsy \cite{liao2013relationship,vaessen2014functional,zhang2011altered}. These studies reveal reduced coupling between tractography-based and functional networks in idiopathic generalized epilepsy, where hub and default mode network connectivity was particularly diminished \cite{zhang2011altered}. In contrast, a study comparing fMRI derived networks with cortical thickness derived networks showed increased coupling between functional and morphometric networks \cite{liao2013relationship}. Thus, the nature of the relationship between anatomical, morphometric, and functional brain networks is a relatively open area in epilepsy. Multimodal anatomical studies combined with computational models may prove useful for pushing forward this frontier.

\paragraph{An Integrative Summary}

In recent years, different investigators have focused on different aspects of network organization within and between diseases, and clinical network neuroscience does not yet provide a complete picture of aberrant network activity. While the application of network neuroscience to clinical disorders is in its infancy, some tentative initial conclusions can be drawn. First, neurological disorders result in detectable anatomical and functional network changes at the global, mesoscale, and regional levels. Second, damage to hub regions, such as those found in the structural rich club or those that participate in communication between many brain systems, is characteristic of several neurological disorders and associated with general failures of cognitive function. Third, more focal damage is associated with more focal cognitive pathology most commonly when it affects local hubs or non-hubs. Finally, neuroimaging techniques sensitive to different spatiotemporal scales of neural function uncover different effects of nervous system disorders on brain network organization and function. In the ongoing effort to describe, understand, and treat nervous system disorders, network science is well-positioned to fundamentally alter our conceptual models and inform translational efforts to enhance quality of life in these patients. 

\paragraph{Caveats and Methodological Challenges}

As this survey attests, concepts from graph theory have provided novel approaches to neuroimaging data collected in health and in nervous system disorders. Nevertheless, the current applications of the tools have important liabilities and limitations. One liability concerns the potential misuse of network techniques as a proxy for theory. While the brain is undoubtedly a network, a truly explanatory network theory of neurological disorders would require an account of how cognitive function occurs and also how it fails in the context of neuropathology. Whereas global, mesoscale, and regional patterns of network failure accompany nervous system disorders, a complete theoretical account that predicts and explains diverse neurological patterns of dysfunction confronting neurologists has not been established. 

Another important consideration concerns the inherent limitations of data collection and analysis. There exists no comprehensive measurement technique at any spatiotemporal scale of neural organization. In addition, neuroimaging data require many processing steps for network construction and interpretation, and these steps differ from one group to another potentially explaining variability in observed network phenotypes \cite{Stam2014}. For future progress, it is imperative that the field work towards a principled and agreed-upon processing framework to accurately and reliably detect network-level effects in nervous system disorders.

\section*{The Future of Network Analysis in Nervous System Disorders}

In this section, we describe a few methodological frontiers and scientific goals in the emerging area of clinical network neuroscience. 

\subsection*{Methodological Frontiers}

\paragraph*{Dynamic Networks}

The brain displays nonstationary dynamics, and an understanding of these dynamics is necessary for an understanding of cognition. Tools to study the brain as a time-evolving networked system include dynamic community detection \cite{Bassett2013}, time-varying independent components analysis \cite{Calhoun2014}, tensor factorization \cite{lee2007nonnegative}, and energy landscape representations \cite{watanabe2014energy}. Efforts in applying these tools to neuroimaging data could help to develop a taxonomy of brain network dynamics in healthy individuals, thereby providing a baseline against which to compare altered brain network dynamics in nervous system disorders.

\paragraph{Multiscale Analysis.}

\begin{figure}[h!]
	\centerline{\includegraphics[width=3.5in]{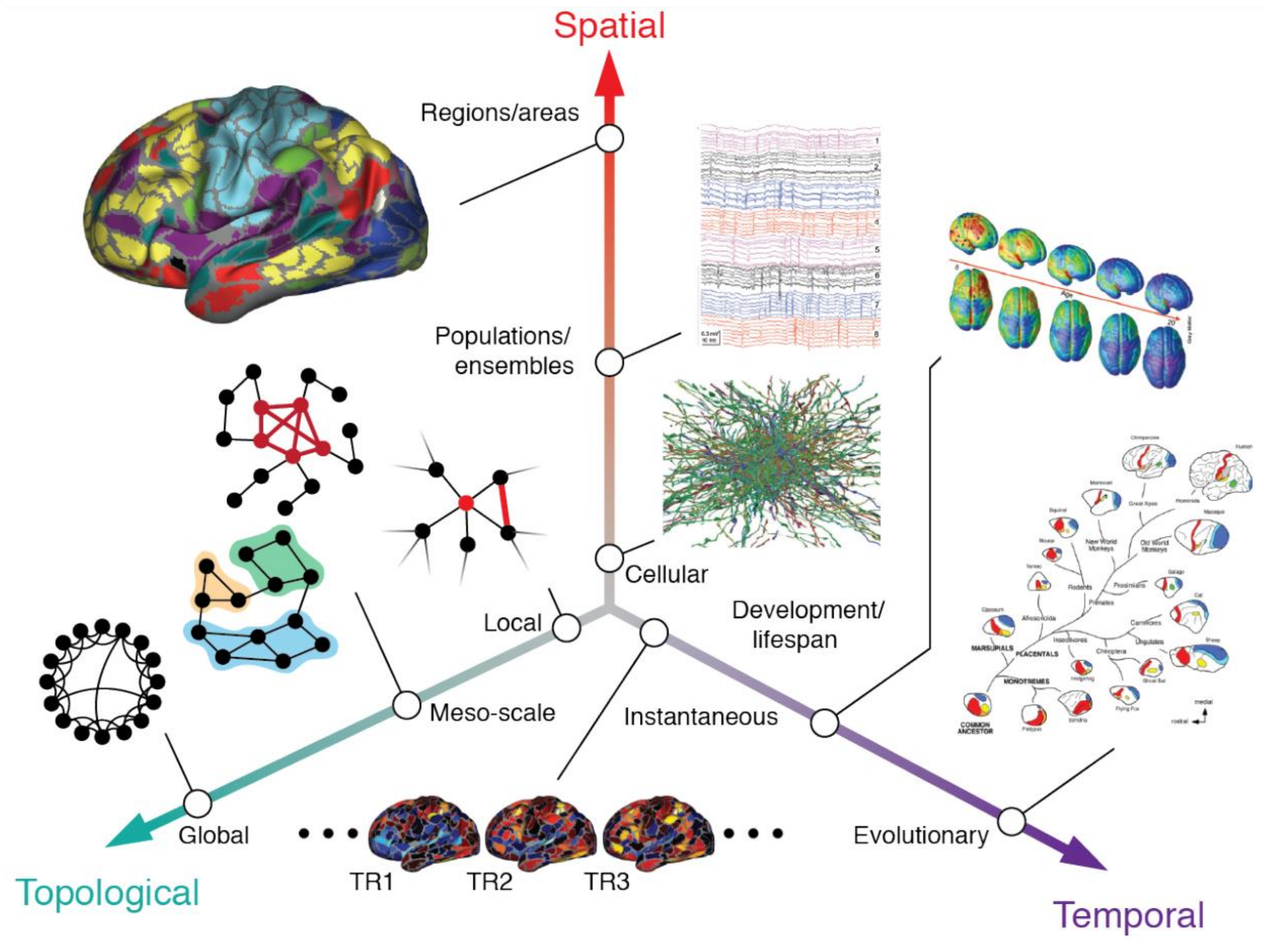}}
	\caption{\textbf{Multiscale network analysis.} Brain networks are organized across multiple spatiotemporal scales and also can be analyzed at topological (network) scales ranging from individual nodes to the network as a whole. TR = repetition time in an fMRI time series. Figure reproduced with permission from \cite{betzel2016multi}.}
	\label{fig:multiscale}
	\centering
\end{figure}

Modern neuroscience provides diverse, increasingly precise, and rapidly accumulating data across many scales of brain organization, from molecules and genes to cell, neuronal ensembles, and large-scale brain areas. How do failures in neurophysiological processes at one level of organization affect another? To address this question, we must have empirical and computational tools to assess the multiscale nature of the brain. There exist many opportunities to extend network neuroscience tools across spatial scales, species, and data types to clarify fundamental principles of neuropathology and neuropsychological symptoms. See \cite{betzel2016multi} for a thorough conceptual overview of multiscale network architecture in human neuroimaging (See Fig. \ref{fig:multiscale}). 

\paragraph*{Anatomy-Function Relationships}

Meaningful interpretations of network alterations in nervous system disorders would benefit from an understanding of how the brain's dynamics are mediated by underlying anatomy. At the scale of non-invasive neuroimaging techniques, inter-regional communication is mediated by axonal bundles. Indeed, the tight relationship between structure and dynamics in the brain motivates ongoing efforts to understand the similarities and differences between anatomical network and functional networks. Studies indicate that anatomical networks (and especially hubs) partially guide functional dynamics in healthy \cite{honey2010can,hermundstad2013structural} and diseased \cite{liao2013relationship,vaessen2014functional,zhang2011altered} human brains. Explicit mathematical models may be critical to fully understand the link between anatomical networks and functional dynamics from a mechanistic perspective. 


\paragraph{Multimodal Predictive Analysis.}

Predicting disease progression and outcomes is an important aspect of clinical and translational research. Efforts in data analysis and modeling have been suggested to neatly separate into two main cultures \cite{breiman2001statistical}. The first culture assumes that a specific process (stochastic or otherwise) produces the data at hand, and therefore uses parametric or nonparametric statistics to explicitly model relationships between independent and dependent variables. The second culture is not as concerned with an explicit model of the data, but instead seeks to use algorithms (e.g., machine learning) to identify patterns in the data, whether or not those patterns provide immediate insight into the underlying mechanisms. Both approaches have proven valuable in understanding network changes in nervous system disorders: the first has provided some basic intuition and targets for treatment, while the second has provided disease predictions that could be used for clinical diagnosis and monitoring. 

\subsection*{Scientific Goals and Clinical Considerations}

\paragraph{Network Mechanisms.}

In this review, we have purposely used the term ``mechanism" as freely as it is applied in the literature at large. Yet, it behooves the reader to ask the question: Exactly what is a mechanism? In one reasonable philosophical account, a mechanism for a phenomenon consists of entities and activities that are responsible for the phenomenon \cite{Illari2012}. This definition maintains that there is a notion of cause and effect (from entities and activities to phenomena) inherent within a system's complex organization that results in phenomena. This same perspective is epistemically useful in network neuroscience as it has been applied to the study of human cognition \cite{medaglia2015cognitive}. However, network mechanisms -- as defined in this fundamental way -- have not yet been rigorously proven in nervous system disorders. 

\paragraph{The Future of Clinical Nosology.}

Clarifying network mechanisms of nervous system disorders could support the refinement of current clinical nosology. Efforts such as the NIH's RDoC explicitly emphasize core principles and dimensions of brain and psychological organization \cite{insel2009endophenotypes}, which could be directly informed by a growing understanding of the network organization that underlies both brain and behavior. Indeed, because network analysis explicitly bridges ``localizationist'' and ``distributed'' theories of brain function \cite{heilman2010clinical}, it can be used to ask questions such as: What are the fundamental relationships between network development and damage that predict cognitive function? What structure-function network alterations underlie the diverse clinical pathology observed in nervous system disorders? If these questions can be answered from a fundamentally network perspective, it suggests the presence of network-level mechanisms governing neurological presentations. 

\paragraph{Clinical Validation.}

Independent of mechanism, an essential question confronts clinical network neuroscientists: how can network analysis provide benefits in the clinic? The clinical imperative requires robust techniques that inform clinical diagnostics, prediction, and treatments \cite{ioannidis2016most}. The goal is to identify clinically important network features by examining whether those features have high diagnostic validity, predict a substantial portion of cognitive variability and change over time, or identify mechanisms of nervous system disorders, neuroplasticity, and recovery. While data to date support the potential utility of network approaches for these goals, currently network statistics are not yet used in the clinic. Future progress will require explicit randomized control trials to prove utility above the current benchmarks.

\paragraph{Cognitive Resilience and Vulnerability.}

The onset, rate of change, and degree of pathology observed in neurodegenerative disorders varies considerably across individuals. A notion known as ``reserve'' -- the properties of the brain responsible for resilience to decline -- has been postulated to mediate this resilience \cite{satz1993brain}. While some environmental and brain factors related to reserve have been identified, a formal network theory for brain reserve does not yet exist. Explaining cognitive resilience and selective vulnerability \cite{seeley2008selective} in nervous system disorders is at a premium when identifying risk factors and designing clinical support. Examining network organization as a mediator for variable trajectories of cognitive decline will form an important goal for the clinical network neuroscientist. 

\paragraph{Personalized Clinical Targeting.}

How can network analysis guide clinical interventions? It is widely known that clinical outcomes are hard to predict, underscoring our limited understanding. Network techniques hold promise by identifying personalized ``fingerprints" for each individual \cite{finn2015functional}. Many clinical interventions offer limited access to disease mechanisms and often with only partial knowledge of these mechanisms. As network neuroscience develops, it may be possible to characterize personalized phenotypes and identify the appropriate spatiotemporal scale for therapeutic targeting at the individual level. While at this stage purely speculative, it is possible that this potential may intersect with approaches such as the use of invasive and non-invasive brain stimulation in control engineering paradigms \cite{medaglia2016mind}, pharmacological agents, and hybrid therapies such as ultrasonic ablation used to allow medicine to cross the blood brain barrier at specific sites \cite{mcdannold2006targeted}. 

\paragraph{Network Control.}

Personalized clinical targeting may be supported by the use of control theory. Once a networked system is well-defined, one can use emerging mathematics to identify points in the network at which to intervene to \emph{control} that system \cite{Ruths2014,liu2011controllability}. Neural control strategies have been applied to solve problems in invasive stimulation \cite{Schiff2012}, and explicitly network control strategies \cite{betzel2016optimally} have been postulated as a mechanism of cognitive control specifically \cite{Gu2015} and mental functions more generally \cite{medaglia2016mind}. In clinical contexts, the goal is to identify strategies that guide the brain's observed function back toward a clinically optimal \emph{controlled} functional regime that supports healthy dynamics \cite{Schiff2012,medaglia2016mind}. This is a largely open frontier requiring collaboration between neuroengineering, neurology, and cognitive neuroscience.

\paragraph{The Human: From Genes to Social Groups.}

With such extensive open terrain to explore, it is easy to remain focused on the brain as an organ enclosed in a skull. However, it is critical to note that the brain is part of an \emph{extended phenotype} \cite{dawkins1999extended} that encompasses the brain, its body, and its environment, both in the clinic and in daily life. Notably, the human organism can be conceptualized from a network perspective across its many spatial scales: from genes to social contexts \cite{dhand2016social}. Such a mutliscale approach may be particularly critical for an understanding of the properties governing cognitive resilience and bodily decline. If network science is uniquely well-suited for anything beyond preceding approaches, it is for exactly the kind of complexity evident in multiscale and multimodal genetic, behavioral, brain, and societal data. 

\section*{Conclusion}

In closing, we note that network analysis in nervous system disorders has begun to revolutionize the vocabulary and concepts necessary to understand, predict, and treat neurological syndromes. Tools are rapidly developing and theoretical perspectives are encouraging scientists to work in communities that creatively study and treat a more comprehensive (and necessarily complex) representation of the human brain. As findings and theories emerge, network neuroscience may establish itself as a key foundation of nervous system disorder nosology and interventions.

\clearpage
\newpage

\bibliographystyle{vancouver} 
\bibliography{JDMReferences}

\end{document}